%% file: 000_main.tex
\documentclass[]{spie}  %>>> use for US letter paper
\input{packages}

\title{Thermal Characterization of a 6-Positioner, 6.2-mm-Pitch Module for Stage-5 Telescopes}

\author[a]{Maxime Rombach}
\author[a]{Malak Galal}
\author[a]{Jonathan Wei}
\author[b]{Stefane Caseiro}
\author[b]{Corentin Magnenat}
\author[a]{Jean-Paul Kneib}
\affil[a]{Institute of Physics, Laboratory of Astrophysics, Ecole Polytechnique Federale de Lausanne (EPFL), Observatoire de Sauverny, CH-1290 Versoix, Switzerland}
\affil[b]{Micro Precision Systems AG (MPS), Bienne, Switzerland}

\authorinfo{Further author information: (Send correspondence to M.R. and M.G.)\\M.R.: E-mail: maxime.rombach@epfl.ch, Telephone: +41 21 693 28 76\\  M.G.: E-mail: malak.galal@epfl.ch, Telephone: +41 21 693 92 91}

\begin{document} 
\maketitle

\begin{abstract}
Ensuring thermal stability of robotic fiber positioners is essential for reliable operation in the real environments of Stage-5 telescopes, where temperature variations can influence mechanical behavior and impact fiber-target accuracy. We present the results of thermal qualification tests conducted on 6.2-mm-pitch robotic positioner modules developed for high-density fiber positioning in next-generation astronomical systems. The positioners were characterized at discrete temperatures spanning –20 °C to +30 °C, representative of expected operational conditions. At each temperature point, key performance metrics, positioning repeatability, hard-stop repeatability, backlash, and non-linearity, were measured and compared to nominal performance. Across the full temperature range, the positioners maintained stable behavior with no measurable degradation in any metric and no evidence of mechanical or electrical damage. These results confirm that the 6.2-mm-pitch architecture provides the necessary thermal resilience for deployment in Stage-5 telescope instrumentation.  
\end{abstract}


\input{001_introduction}
\input{002_System_overview}
\input{003_Methodology}

\input{004_Results}

\input{005_Conclusion}

% \acknowledgments % equivalent to \section*{ACKNOWLEDGMENTS}       
 
% This unnumbered section is used to identify those who have aided the authors in understanding or accomplishing the work presented and to acknowledge sources of funding.  

% References
% \bibliography{references} % bibliography data in report.bib
\bibliography{references_FIXED} % bibliography data in report.bib
\bibliographystyle{spiebib} % makes bibtex use spiebib.bst

\end{document}

%% file: packages.tex
\usepackage{amsmath,amsfonts,amssymb}
\usepackage{booktabs}
\usepackage{float}
\usepackage{graphicx}
\usepackage{gensymb}
\usepackage[colorlinks=true, allcolors=blue]{hyperref}
\usepackage{subcaption}

\usepackage{lmodern}
\usepackage{siunitx}
\usepackage{comment}
\usepackage[justification=centering]{caption}
\usepackage{adjustbox}

%% file: 001_introduction.tex
\section{INTRODUCTION}
\label{sec:intro}  % \label{} allows reference to this section

Future Stage-5 astronomical surveys aim to have a high fiber multiplexing with a typical number of 20'000; while the highest multiplexing is currently achived by the DESI experiment with 5'000 optical fibers\cite{silber_robotic_2023}. Positioning such a high number of fibers in parallel requires an equal amount of robotic positioners. Different focal plane concepts are currently being studied to \textit{modularize} the focal plane assembly. This study focuses on testing the first 6-robots prototype of triangular module conceptualize in Silber, et al. (2022)\cite{silber_25000_2022}.

\begin{figure}[H]
     \centering
     \begin{subfigure}[b]{0.49\textwidth}
        \centering
        \includegraphics[width=0.6\linewidth]{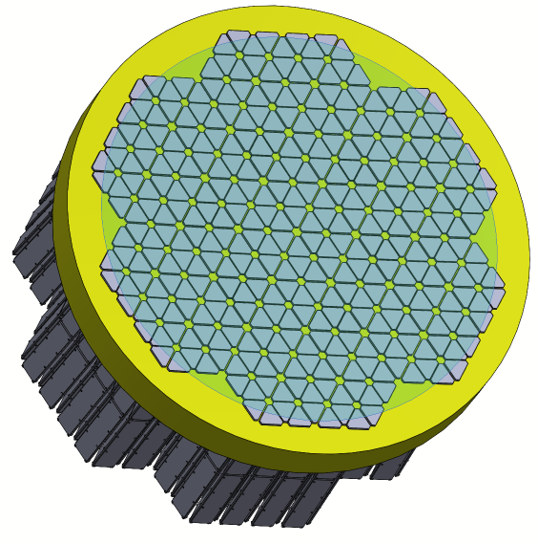}
        \caption{Focal plane modular architecture with triangular modules of positioners ($\approx$ 180 to 460 modules depending on the instrument project)}
        \label{fig:MM_FP}
     \end{subfigure}
     \hfill
     \begin{subfigure}[b]{0.49\textwidth}
         \centering
         \includegraphics[width=0.8\textwidth]{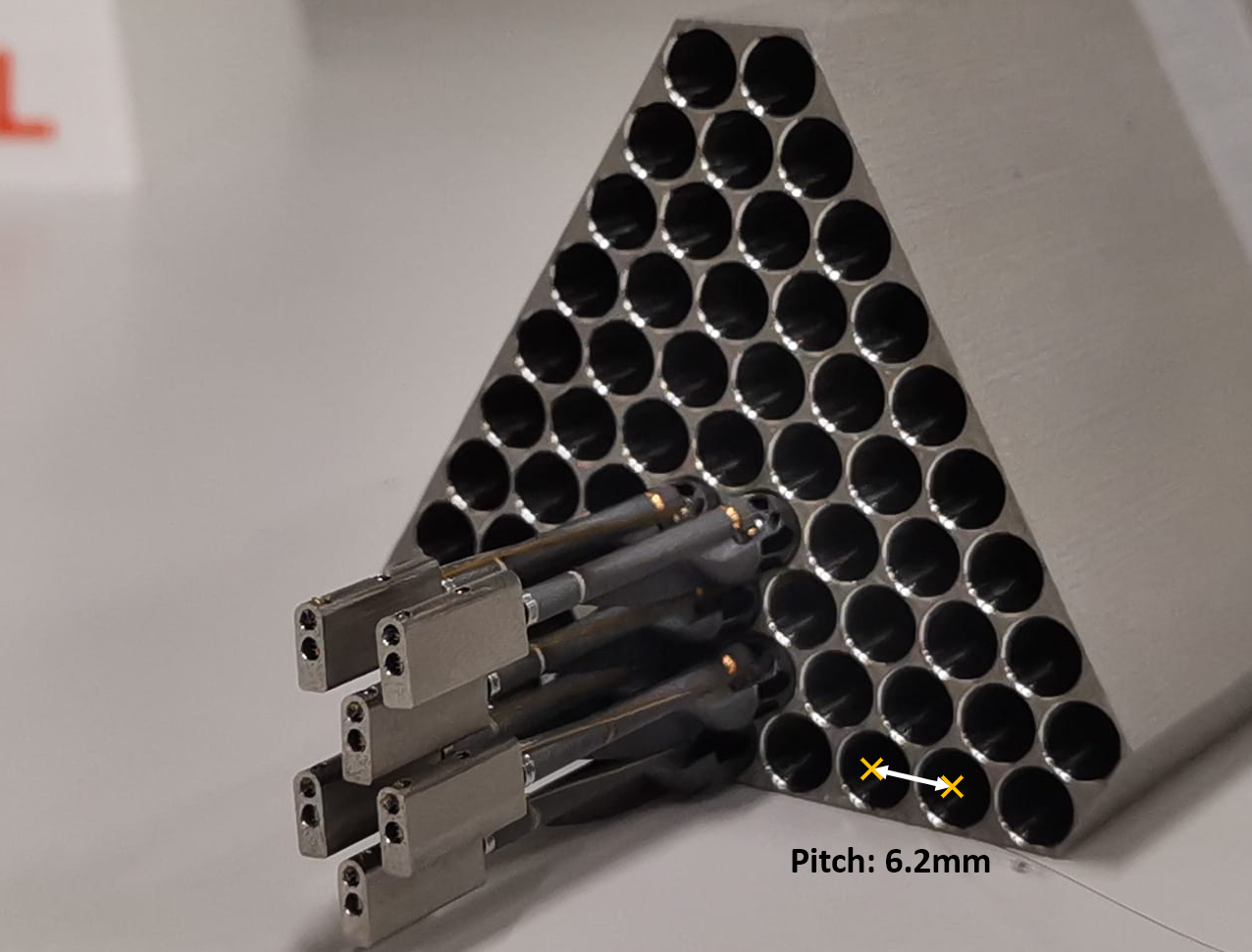}
         \caption{MPS6 prototype: 6 fiber positioners packaged in a chassis planned to host 63 positioners with 6.2mm pitch}
         \label{fig:MPS6_pitch}
     \end{subfigure}
        \caption{Modular focal plane concept and first module prototype}
        \label{fig:Intro_images}
\end{figure}
\subsection{Objectives}
A first test campaign under lab conditions, proved that this modular approach with 6.2mm-pitch positioners modules can perform according to specifications \cite{galal_prototyping_2026}. Since the positioners are likely to experience a wide temperature range we want to check if their performances remain stable with temperature change. The module prototype tested here is the same as in Galal, et al. (2026)\cite{galal_prototyping_2026} manufactured by Micro Precision Systems (MPS), Biel - Switzerland. It will be later referred as "MPS6", see Figure \ref{fig:MPS6_pitch}.\\
The measured metrics are the same as well namely:

\begin{itemize}
    \item \textbf{Datum repeatability:} the robots' ability to consistently stop at the same location upon reaching the arms hardstops
    \item \textbf{Positional repeatability:} the robots' ability to return to the same position for a given command
    \item \textbf{Backlash:} mechanical clearance between coupled gears in the system, leading to uncertainty in the motion of the arms
    \item \textbf{Non-linearity:} local change of the gear ratio due to gears manufacturing and assembly uncertainty, leading to uncertainty of positioning, particularly seen in small angular motion (typically 1 deg) 
\end{itemize}

The robots are usually operated atop mountains which are affected by thermally straining environments. The above characteristics will be evaluated throughout a typical operational range of the robots on the telescope: -20°C, -10°C, 5°C, 20°C, 30°C.

%% file: 002_System_overview.tex
\section{System overview}
\begin{figure}[H]
     \centering
     
     % --- SUBFIGURE A ---
     \begin{subfigure}[b]{0.32\textwidth}
        \centering
        % This cuts exactly 28% off the left of the actual image safely
        \adjustimage{height=4.8cm, trim={0.22\width} 0 0 0, clip}{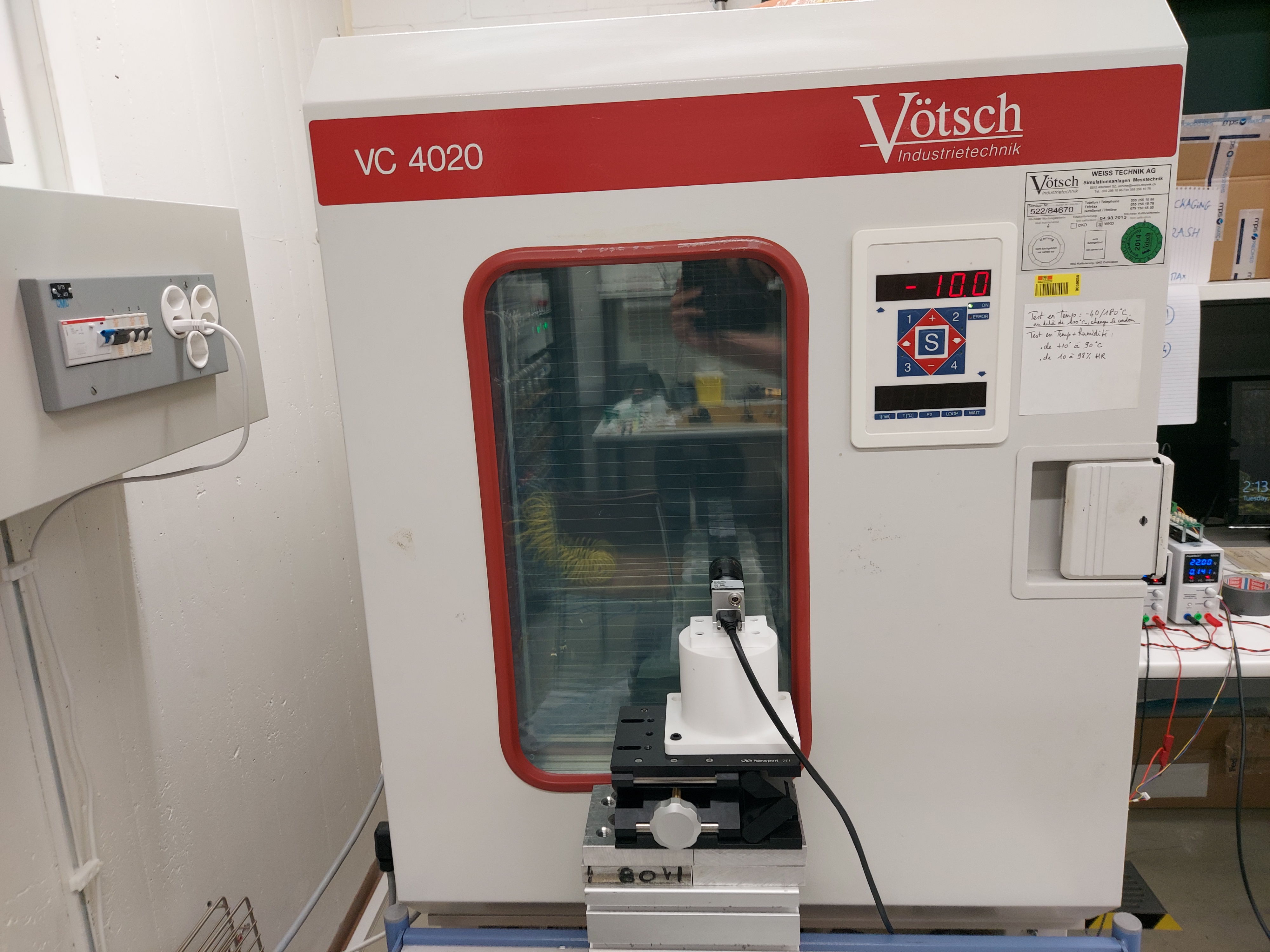}
        \caption{}
        \label{fig:002_camera_closed_door}
     \end{subfigure}
     \hfill
     % --- SUBFIGURE B ---
     \begin{subfigure}[b]{0.32\textwidth}
         \centering
         \includegraphics[height=4.8cm]{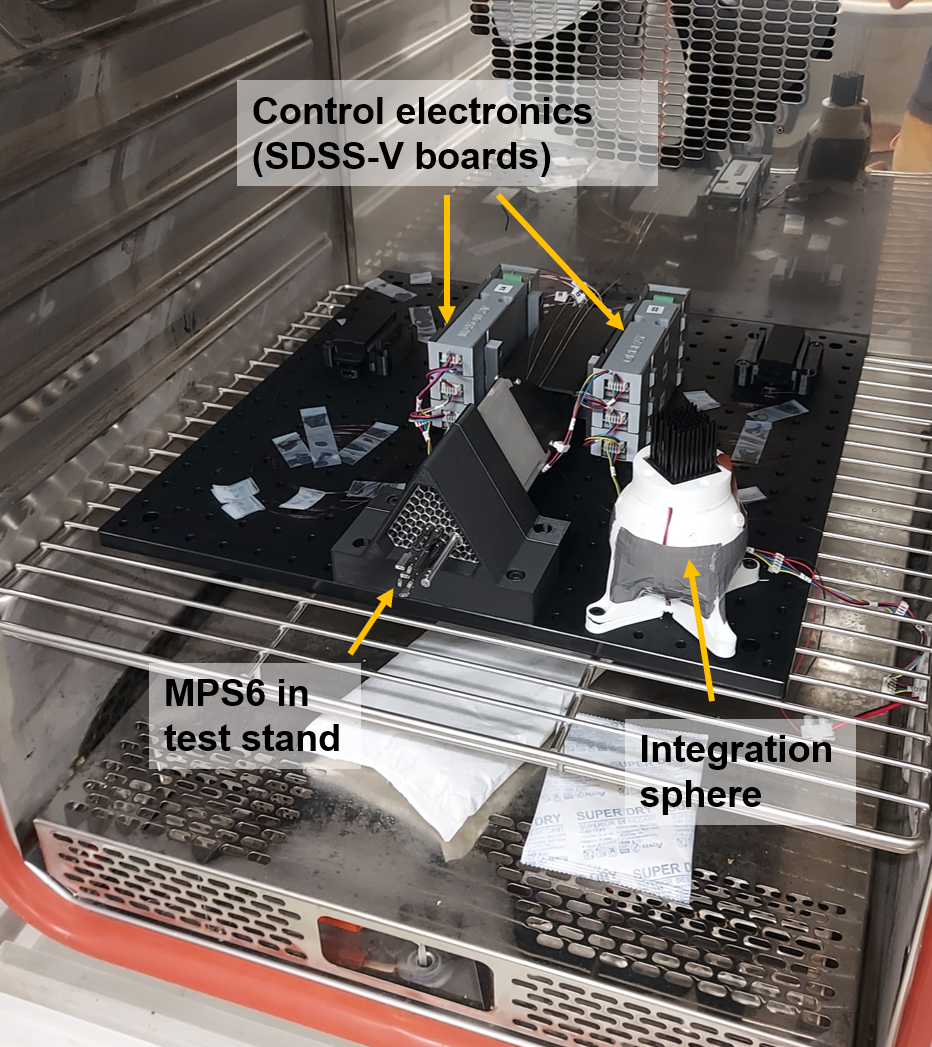}
         \caption{}
         \label{fig:002_pos_in_chb_labelled}
     \end{subfigure}
     \hfill
     % --- SUBFIGURE C ---
     \begin{subfigure}[b]{0.32\textwidth}
         \centering
         \includegraphics[height=4.8cm]{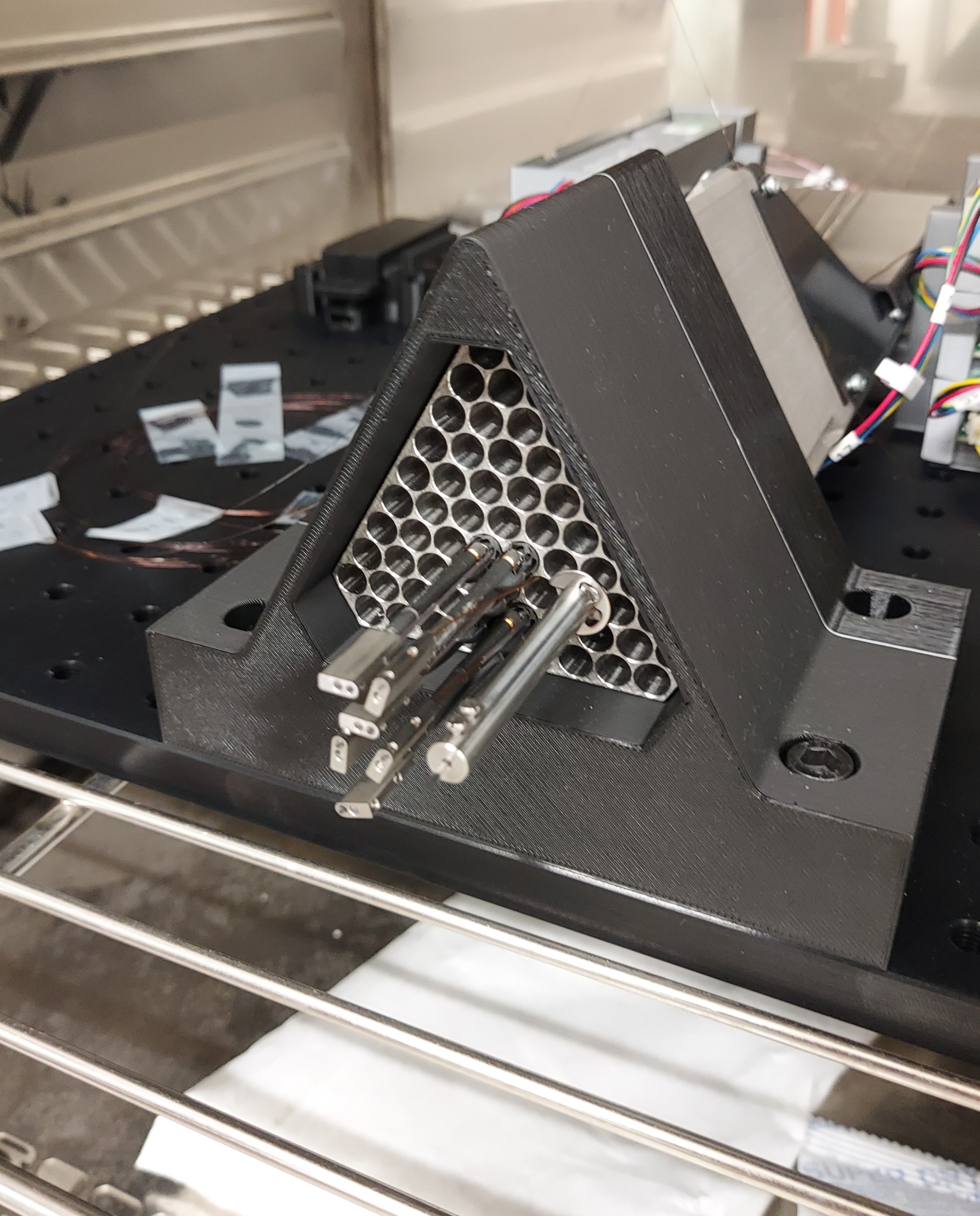}
         \caption{}
         \label{fig:002_pos_in_chb_zoomed}
     \end{subfigure}
     
     % --- MAIN LEFT-ALIGNED CAPTION ---
     \caption{Overview of the thermal test setup; (a) External view of the thermal chamber window showing the camera alignment; (b) Breadboard with mounted positioners and integration spheres inside the thermal chamber; and (c) Zoomed view on the module prototype bolted in the test stand.}
     \label{fig:002_thm_chb_overview}
\end{figure}

The fiber positioner module prototype of 6 robots is bolted on a 3D printed black stand in turn secured on a Thorlabs MB4545/M breadboard. Each robot is controlled by an SDSS-V card as previously performed for lab-condition xy tests\cite{galal_prototyping_2026}. The EPFL Astrobots team designed and built this control board for the eponym project, which also used Brushless DC (BLDC) motors for their positioners. They are, hence, used for prototyping purpose while a dedicated electronics is being designed in parallel. The current robots prototypes are actuated by 4mm BLDC motors. \\
A 3D printed integration sphere is also fixed on the breadboard to provide back-illumination of each fiber. Each robot carries an optical fiber on its end effector. They are back-illuminated one after the other at each termination of thermal sequence.
The whole breadboard-assembly is placed inside a Vötsch VC 4020 thermal enclosure inside which temperature can be controlled in a range of -40°C to 180°C, as pictured in Figure \ref{fig:002_thm_chb_overview}\\
A basler acA5472-17um mounted with a HF3520-12M objective lens is positioned outside the thermal chamber close enough to its window to rule out unwanted reflections. The lens is close but non-touching to avoid transmitting the chamber vibrations to the acquisition device.
\begin{comment}
\begin{figure}[H]
     \centering
     \begin{subfigure}[b]{0.33\textwidth}
        \centering
        \includegraphics[width=0.6\linewidth]{figures/002_system_overivew/Camera_closed_door.jpg}
        \caption{}
        \label{fig:002_camera_closed_door}
     \end{subfigure}
     \hfill
     \begin{subfigure}[b]{0.33\textwidth}
         \centering
         \includegraphics[width=0.5\textwidth]{figures/002_system_overivew/setup_in_chamber_labelled.png}
         \caption{}
         \label{fig:002_pos_in_chb_labelled}
     \end{subfigure}
     %
          \begin{subfigure}[b]{0.33\textwidth}
         \centering
         \includegraphics[width=0.5\textwidth]{figures/002_system_overivew/20250616_164046_zoomed2.jpg}
         \caption{}
         \label{fig:002_pos_in_chb_zoomed}
     \end{subfigure}
        \caption{Overview of the thermal test setup; (a) Camera looking at the positioners through the chamber window; (b) Breadboard with mounted positioners and integration spheres inside the thermal chamber; and (c) Zoomed view on the module prototype bolted in the test stand.}
        \label{fig:002_thm_chb_overview}
\end{figure}
\end{comment}

\begin{figure}[!t]
     \centering
     
     % --- SUBFIGURE A ---
     \begin{subfigure}[b]{0.32\textwidth}
        \centering
        \includegraphics[height=4.5cm]{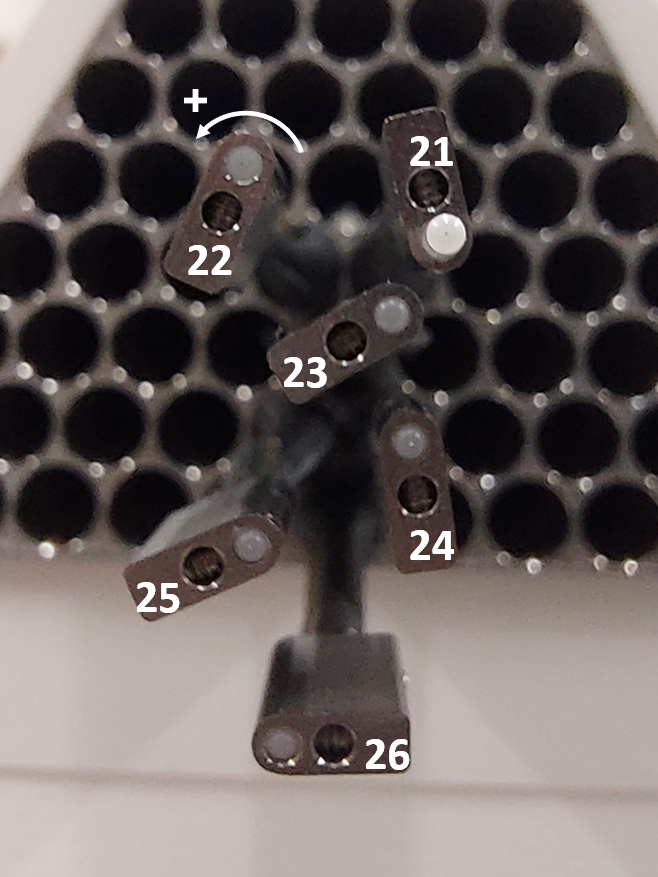}
        \caption{}
        \label{fig:002_labelled_pos}
     \end{subfigure}
     \hfill
     % --- SUBFIGURE B ---
     \begin{subfigure}[b]{0.32\textwidth}
         \centering
         \includegraphics[height=4.5cm]{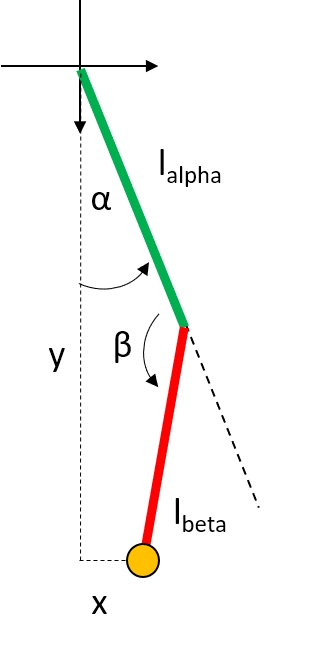}
         \caption{}
         \label{fig:002_pos_schematics}
     \end{subfigure}
     \hfill
     % --- SUBFIGURE C ---
     \begin{subfigure}[b]{0.32\textwidth}
         \centering
         \includegraphics[height=4.5cm]{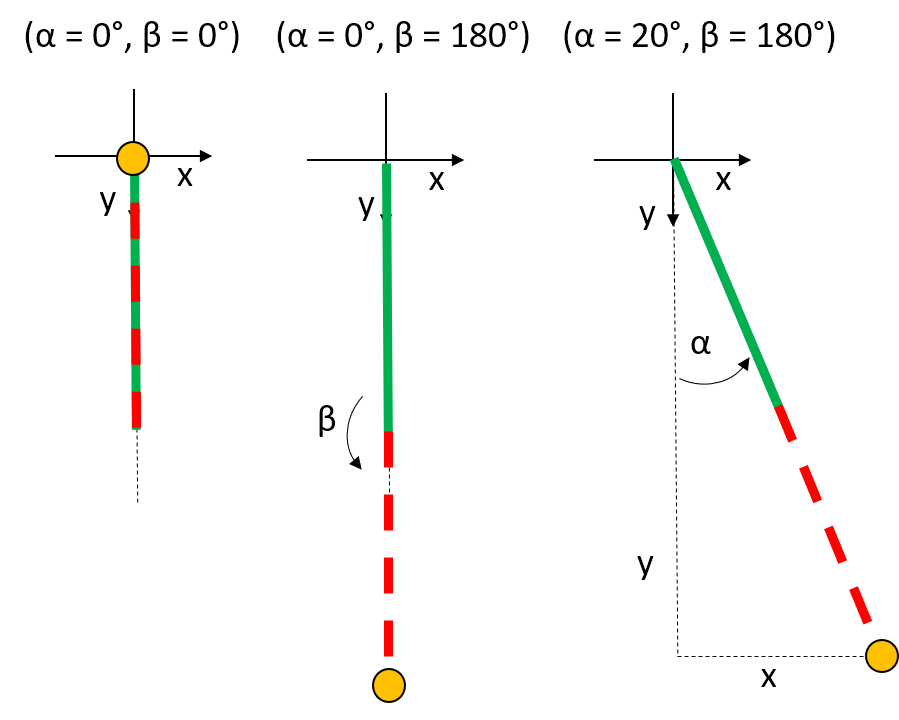}
         \caption{}
         \label{fig:002_pos_schematics_examples}
     \end{subfigure}
     
     % --- MAIN LEFT-ALIGNED CAPTION ---
     \caption{Identification of the positioners and their kinematics; (a) Labeled positioners and convention for positive rotation direction (counterclockwise); (b) Schematic of the robots positioners kinematics; in yellow the optical fiber; and (c) Schematic examples of different angular positions of the positioners.}
     \label{fig:002_pos_overview}
\end{figure}

\noindent The MPS6 has 6 \textit{theta-phi}, or \textit{alpha-beta} positioners identified as pos21, pos22, pos23, pos24, pos25, pos26, each featuring an alpha and beta arm  as shown in Figure \ref{fig:002_pos_overview}.\\
Figure \ref{fig:002_labelled_pos} shows a close-up view on the beta arms of the robots carrying the optical fibers, easily identified here by their 1.25mm ferrule. This ferrule diameter will be reduced in the future to optimize the beta arms footprint and avoid collisions.\\
Schematics of the positioners kinematics in Figure \ref{fig:002_pos_schematics} and \ref{fig:002_pos_schematics_examples} also highlight how each robot moves and draws attention to the beta angle \textit{being relative to the alpha arm}.\\
The range of motion for each axis is [-5°;365°] for alpha and [0°;180°] for beta.

\begin{comment}
\begin{figure}[H]
     \centering
     \begin{subfigure}[b]{0.3\textwidth}
        \centering
        \includegraphics[width=0.8\linewidth]{figures/002_system_overivew/labelled_pos.png}
        \caption{}
        \label{fig:002_labelled_pos}
     \end{subfigure}
    \hfill
     \begin{subfigure}[b]{0.3\textwidth}
         \centering
         \includegraphics[width=0.6\linewidth]{figures/002_system_overivew/schametics_theta_phi.png}
         \caption{}
         \label{fig:002_pos_schematics}
     \end{subfigure}
    \hfill
    \begin{subfigure}[b]{0.3\textwidth}
         \centering
         \includegraphics[width=1.2\linewidth]{figures/002_system_overivew/schametics_theta_phi_pos_examples.png}
         \caption{}
         \label{fig:002_pos_schematics_examples}
     \end{subfigure}
        \caption{Identification of the positioners and their kinematics; (a) Labeled positioners and convention for positive rotation direction (counterclockwise); (b) Schematic of the robots positioners kinematics; in yellow the optical fiber; and (c) Schematic examples of different angular positions of the positioners.}
        \label{fig:002_pos_overview}
\end{figure}
\end{comment}

%% file: 003_Methodology.tex
\section{Testing Methodology}
\subsection{Chamber control}
\label{sec:chamber_control}
One positioner at a time was tested for a full temperature range starting from 30°C and descending temperatures until -20°C. The chamber communicates to the PC via RS232 and a custom python script gives the user a Graphic User Interface (GUI) to easily interact with it.
\begin{figure}[H]
    \centering
    \includegraphics[width=0.6\linewidth]{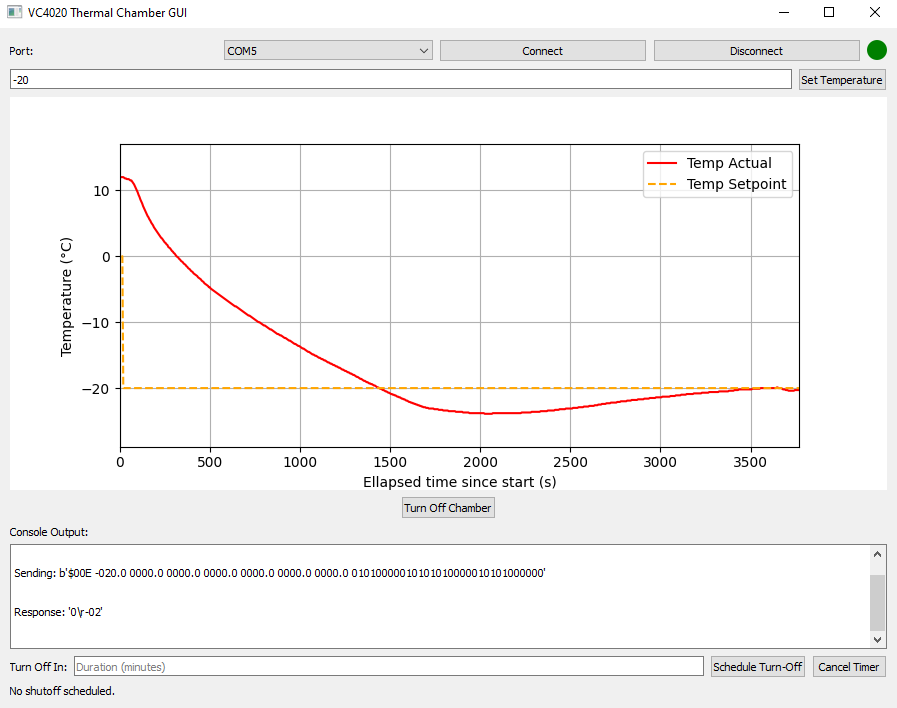}
    \caption{Screenshot of the GUI display; highlighting the settling time of the thermal chamber}
    \label{fig:003_GUI}
\end{figure}
\noindent The temperature displayed corresponds to the native temperature probe inside the thermal chamber. Its exactness has been checked at room temperature with an external temperature probe that we have in the lab. For the current test campaign, we rely solely on the chamber's probe data for this test. For future thermal test campaigns, we plan to use more thermal probes directly on the module itself for a more precise assessment of its temperature. In the meantime we rely on the following thermal assumptions:
\begin{enumerate}
    \item \textbf{Initial temperature settling time}: Figure \ref{fig:003_GUI} shows the chamber takes about hour to hit the temperature setpoint after a slight overshoot for a temperature delta of about -30°C. Accounting for a additional margin for settling afterwards it is assumed that \textit{everything inside the chamber is stabilized at the displayed temperature after 2 hours}.
    \item \textbf{Temperature stability when chamber is off}: the temperature displayed increases up to 2°C maximum for 30 minutes. By overshooting the test temperature by 1°C and assuming the thermal inertia of the module is sufficiently high, \textit{the module’s temperature is then considered to be contained within ±1°C around the nominal test temperature for 30 min when the chamber is off}
    \item \textbf{Internal convection when chamber turned off}: \textit{the air inside the chamber is considered still} when the chamber is turned off, as the fan does not run anymore.
\end{enumerate}

\subsection{Chamber vibrations}
\begin{figure}[H]
     \centering
        \centering
        \includegraphics[width=0.4\linewidth]{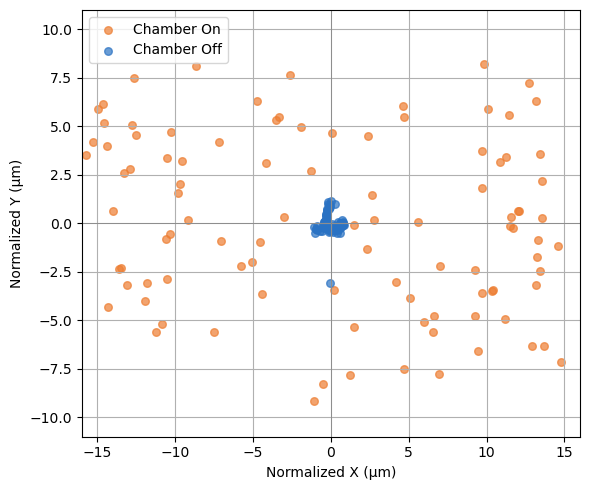}
        \caption{Effect of the vibrations of the climatic chamber on the spread of the data points: measurements at 20° using positioner 22}
        \label{fig:003_vibrations}
\end{figure}

The assumption 2) and 3) discussed above take root in tackling the issue of the chamber vibrations when the temperature control is active; namely when the chamber is turned on.\\
For testing time optimization reasons, the initial assumption on the thermal chamber was that its vibrations were not impacting the measurements too much. To verify this assumption, vibration assessments were realized for scenarios where the chamber was turned on and off. As shown in Figure \ref{fig:003_vibrations}, We captured 100 centroid measurements while not moving the positioners at all, not powered, pos22, with a time step of 0.2s.\\
The below graphs show the centered data around their mean for 20°C and 5°C. They exhibit a noticeable difference in favor of the \textit{chamber off case}.

\noindent While results for test cycles of positioners 24 to 26 were not thrown off despite being run with chamber on, it was decided to conduct the test under assumption 2) and 3) of Sec. \ref{sec:chamber_control} namely:
\begin{itemize}
    \item Chamber is turned off during the measurements acquisition to eliminate the effect of vibrations on each test cycle (defined in Sec \ref{sec:test_cycle})
    \item Assuming the thermal inertia of the module is sufficiently large to remain within ±1°C of the nominal temperature during one test cycle of 30 min
\end{itemize}

\subsection{Test protocol}

\begin{enumerate}
    \item \textbf{Setting-up positioner}: each positioner carries an optical fiber that is backlit one-by-one according to the positioner to be tested. For representation, Figure \ref{fig:003_6_pos_light} shows the point of view of the camera when all 6 fibers are backlit. Figure \ref{fig:003_1_pos_light} then shows a similar with but with only pos24 backlit, which is a typical view of the camera during a test cycle. 
    The lightspot of the fibers is the brightest element in the camera Field Of View (FOV) so we can isolate it easily by lowering the exposure time of the camera so that only the lightspot is left. Image processing is then done to extract the lightspot position in the image. Figure \ref{fig:003_1_pos_light_gaussian} highlights the Gaussian distribution of the light spot.\\

\begin{figure}[H]
     \centering
     
     % --- SUBFIGURE A ---
     \begin{subfigure}[b]{0.32\textwidth}
        \centering
        \includegraphics[height=5.2cm]{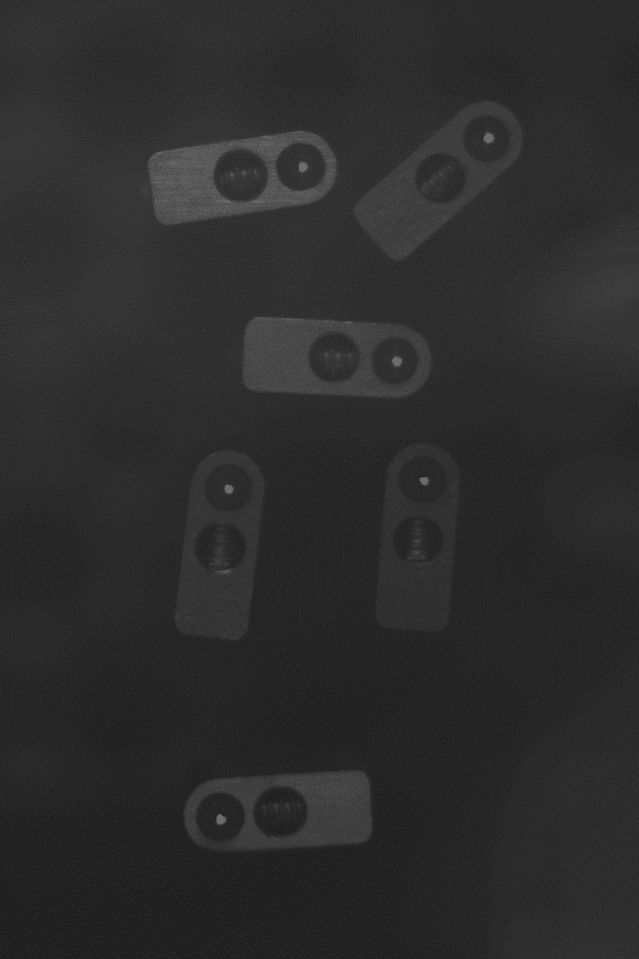}
        \caption{}
        \label{fig:003_6_pos_light}
     \end{subfigure}
     \hfill
     % --- SUBFIGURE B ---
     \begin{subfigure}[b]{0.32\textwidth}
         \centering
         \includegraphics[height=5.2cm]{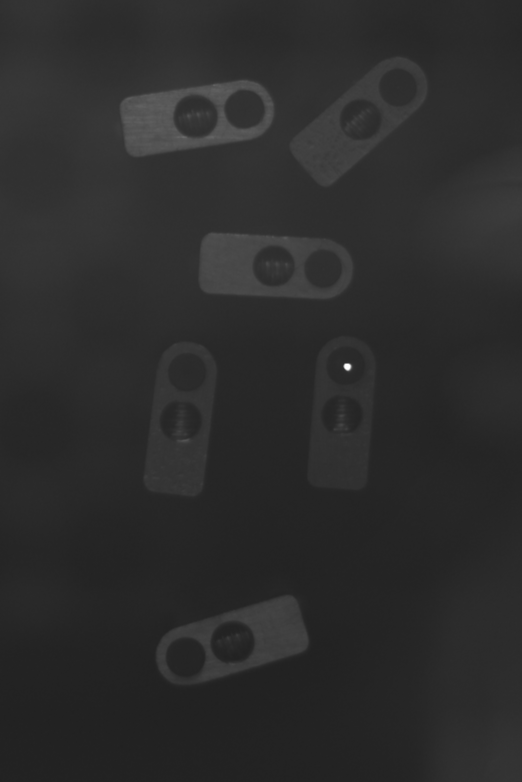}
         \caption{}
         \label{fig:003_1_pos_light}
     \end{subfigure}
     \hfill
     % --- SUBFIGURE C ---
     \begin{subfigure}[b]{0.32\textwidth}
         \centering
         % Trims 1.2cm from the left and 1.2cm from the right side of the image
         \includegraphics[height=5.2cm, trim=1.2cm 0 1.2cm 0, clip]{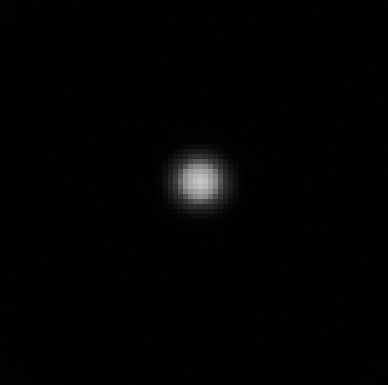}
         \caption{}
         \label{fig:003_1_pos_light_gaussian}
     \end{subfigure}
     
     % --- MAIN LEFT-ALIGNED CAPTION ---
     \caption{Pictures highlight the camera POV through the window of the thermal chamber; (a) All 6 fibers back-lit; (b) Only one back-lit fiber; and (c) Close-up highlight of the Gaussian intensity distribution of a single light spot.}
     \label{fig:003_pos_lightspots}
\end{figure}

    \item   \textbf{Temperature settling}: once the desired fiber is illuminated the chamber control is set to the desired temperature. We wait for the displayed temperature to settle around the desired setpoint for 2 hours before launching a test cycle, according to the assumptions made in Sec. \ref{sec:chamber_control}.
\label{sec:test_cycle}
\item \textbf{Test cycle}: (duration: 30 min) a test cycle is a set of positioner moves after which each position is stored in a results file for later post processing. The test cycle sequences are composed of the following:

    \begin{enumerate}
        \item \textit{Homing}: Move both alpha and beta axes to hardstops; we always start from the home positions; defined as -5° for alpha and 0° for beta according to the convention set in Figure \ref{fig:002_pos_overview}.
        \item \textit{Beta arcs to get the alpha circle}:
    For each alpha position [0, 60, 120, 180, 240, 300], we sweep the beta arm across these [0, 30, 60, 90, 120, 150]. Finding the center of each beta sweep allows us to construct the alpha circle. The latter provides us the position of the center of the positioner and the length of the alpha arm. The radius of a beta sweep corresponds to the beta arm length.\\
    For this first thermal test campaign, we do not take into account the effect that \textit{positioners tilt}, as defined in Malak, et al. (2026)\cite{galal_prototyping_2026}, may have on the measurements. 
    \begin{figure}[H]
        \centering
        \includegraphics[width=0.6\linewidth]{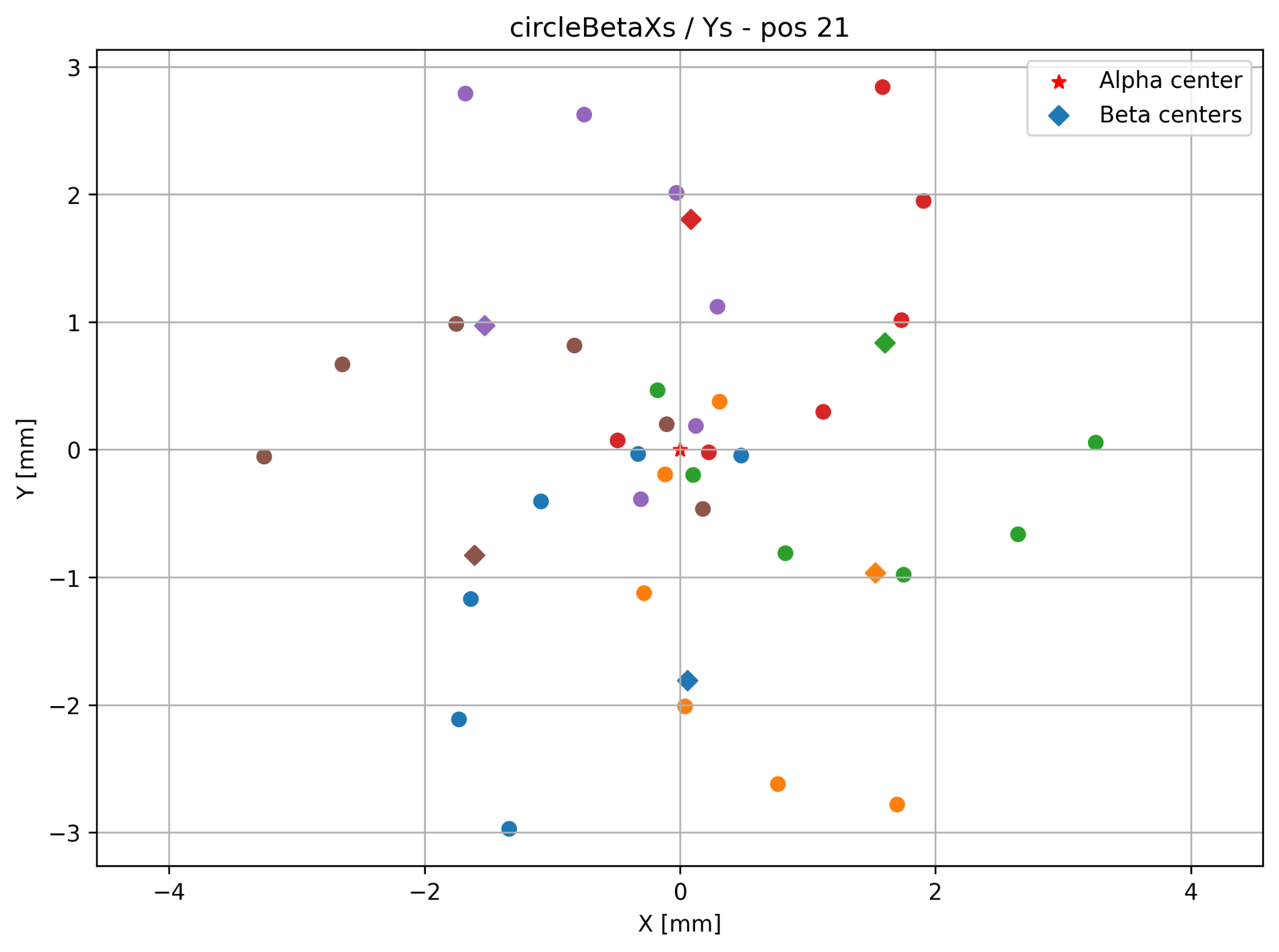}
        \caption{pos21 at 20°C - beta sweeps at different alpha angle to get alpha center and arms lengths}
        \label{fig:003_beta_sweeps}
    \end{figure}
    
    \item \textit{Datum (or hardstops) repeatability}: test how repeatable is the fixed reference of both arms.
    One arm at a time, the tested arm is moved 20° away from the hardstop position and moved back relatively 30° to overcome backlash and ensure going back against the physical hardstop. Each time an arm’s hardstop is tested, and the other is set to its 0° position. For this test campaign, 10 iterations were conducted to optimize test cycle time.
    
    \item \textit{Positional repeatability}: ability of each arm to consistently perform the same movement for the same angular command. The test procedure is the same for both arms. The tests parameters are the following:
        \begin{itemize}
            \item While testing alpha, the beta arm is set to 180°
            \item While testing beta, the alpha arm is fixed to 0°.
            \item Test position: 30°
            \item Delta move: 10°
        \end{itemize}
    The move sequence is then as follows:
        \begin{enumerate}
            \item The arm is sent to the test position + delta move
            \item The arm travels back to the test position. The direction change implies that the arm has to overcome its internal backlash before moving. The centroid of the fiber coming from a higher angular position is then acquired, thus generating the alpha (or beta) \textit{high} data set.
            \item The arm then move to the test position - delta move
            \item The arm travels back to the test position. Similarly from ii) the backlash needs to be overcome upon direction change. This time the arm approaches the test position from a lower angular position, thus giving the alpha (or beta) \textit{low} data set.
        \end{enumerate}
    
    \noindent Once the 2 datasets are obtained the repeatability of each arm is calculated from each high dataset.

    \item \textit{Backlash}: backlash for both arms is deduced from the angular difference of their \textit{High} and \textit{low} dataset.

    \item \textit{Non-linearities}: Data for non-linearities are taken by moving each axis by steps of 1° from their hardstops across the whole angular range. As a result, this is the most time consuming parameter to test in a test cycle. When testing the alpha arm, beta is fixed to the 180° position, its fully extended position, see Figure \ref{fig:002_pos_schematics_examples}.

\end{enumerate}

The test sequences in c), d), e) and f) are thoroughly discussed in Galal et al. (2026)\cite{galal_prototyping_2026}.

Each test cycle takes 30 min to run during which the chamber is off as per the reasons previously mentioned. After each test cycle the chamber is turned on again to lower down the temperature to the desired setpoint. For test time length reasons, we also decided that running 3 test cycles per positioner and temperature was acceptable.

\end{enumerate}

%% file: 004_Results.tex
\section{Experimental results}
\label{sec:004_results}
The tests results presented below are obtained via the root mean square (RMS) of the parameters discussed above over each 3 test cycles per positioner and temperature.
\subsection{Datum repeatability}
\begin{figure}[H]
     \centering
     \begin{subfigure}[b]{0.49\textwidth}
        \centering
        \includegraphics[width=0.8\linewidth]{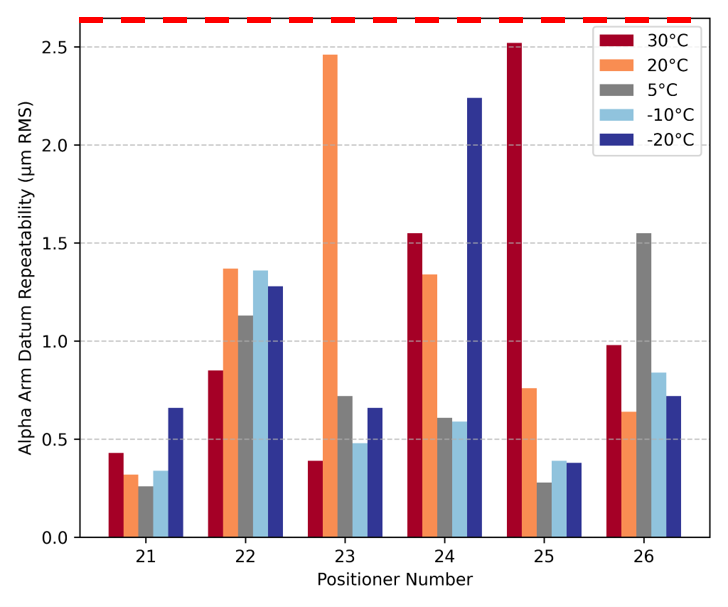}
        \caption{Alpha arm}
        \label{fig:004_dat_a_arm}
     \end{subfigure}
     \hfill
     \begin{subfigure}[b]{0.49\textwidth}
         \centering
         \includegraphics[width=0.8\textwidth]{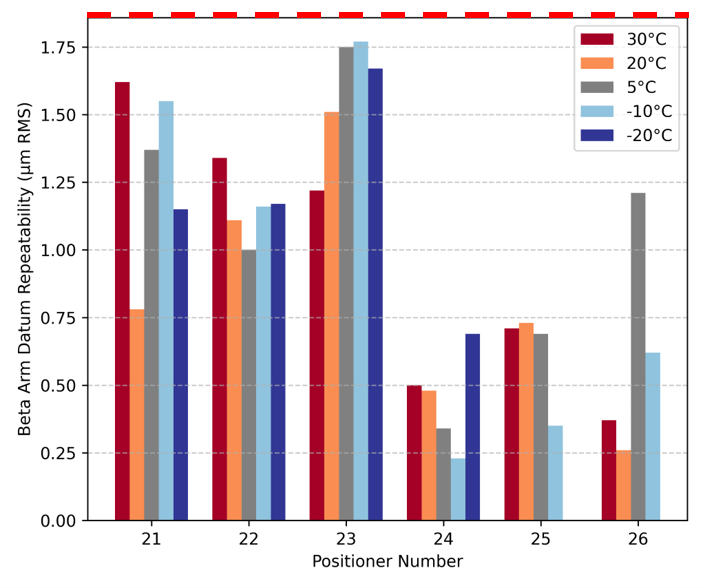}
         \caption{Beta arm}
         \label{fig:004_dat_b_arm}
     \end{subfigure}
        \caption{Graphs showing the datum repeatability for MPS6. The bar plots show the root-mean-square (RMS) values versus the positioner number (Specification: 50 $\mu m$)}
        \label{fig:004_dat}
\end{figure}

Figure \ref{fig:004_dat} presents the results of the datum repeatability that show consistency in the data with values below 2.5 µm RMS. However, a strange behavior of positioners 25 and 26 at -20°C was observed when calculating the datum repeatability for the beta arm, and the values of repeatability were extremely high indicating an problem for these two positioners. 

\subsection{Positional repeatability}
\begin{figure}[H]
     \centering
     \begin{subfigure}[b]{0.49\textwidth}
        \centering
        \includegraphics[width=0.8\linewidth]{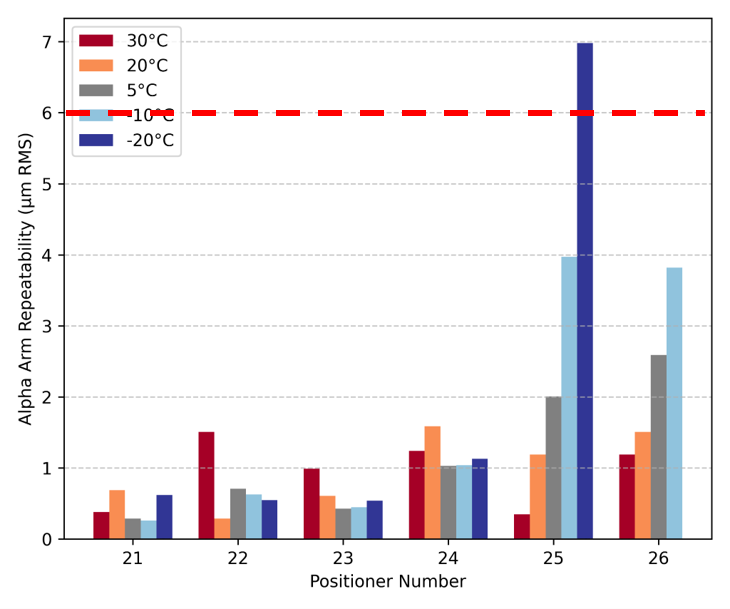}
        \caption{Alpha arm}
        \label{fig:004_rep_a_arm}
     \end{subfigure}
     \hfill
     \begin{subfigure}[b]{0.49\textwidth}
         \centering
         \includegraphics[width=0.8\textwidth]{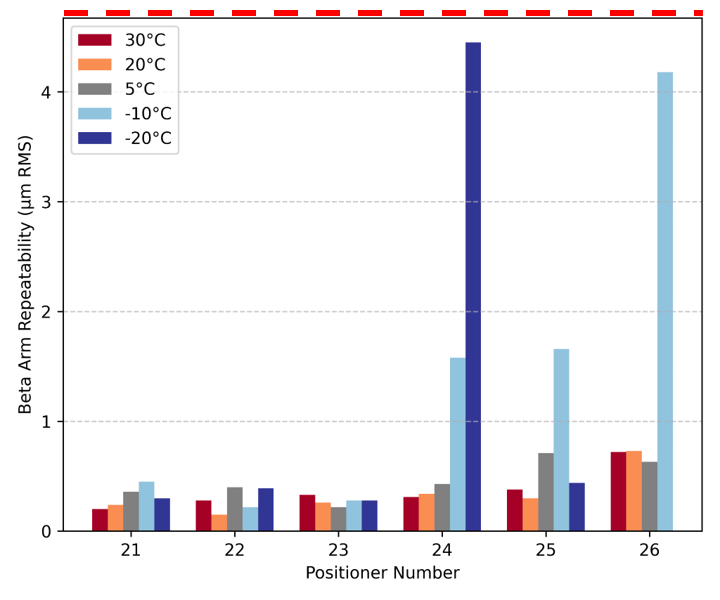}
         \caption{Beta arm}
         \label{fig:004_rep_b_arm}
     \end{subfigure}
        \caption{Graphs showing the repeatability for MPS6. The bar plots show the root-mean-square (RMS) values versus the positioner number (Specification: 6 $\mu m$)}
        \label{fig:004_rep}
\end{figure}

The results presented in Figure \ref{fig:004_rep} show the XY positioning repeatability of both alpha and beta arms. As can be seen in the figure, the repeatability of both arms exhibits a uniform behaviour with values below 3 µm RMS with few anomalies at the negative temperatures (-10°C and -20°C) for some positioners (24 and 25) reaching values up to 7 µm RMS for the alpha arm and 5 µm RMS for the beta arm. Positioner 26, however, showed a strange performance at low temperatures. No result bar can be seen for both arms of pos26 at -20°C. That indicates a high probability of the positioner being immobile during the test cycle at this particular temperature. Test cycles were ran multiple times at this temperature all exhibiting the same results. We even rose the temperature back to -10°C to test if the positioner was still able to move and obtained similar results as for previous tests at this temperature. Back to 20°C the positioner was still exhibiting the same "immobile" behavior.

\subsection{Backlash}
\begin{figure}[H]
     \centering
     \begin{subfigure}[b]{0.49\textwidth}
        \centering
        \includegraphics[width=0.8\linewidth]{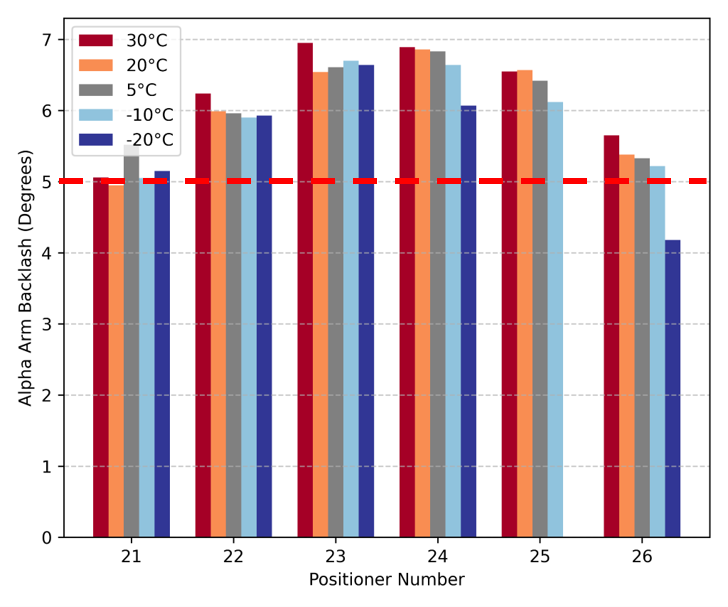}
        \caption{Alpha arm}
        \label{fig:004_back_a_arm}
     \end{subfigure}
     \hfill
     \begin{subfigure}[b]{0.49\textwidth}
         \centering
         \includegraphics[width=0.8\textwidth]{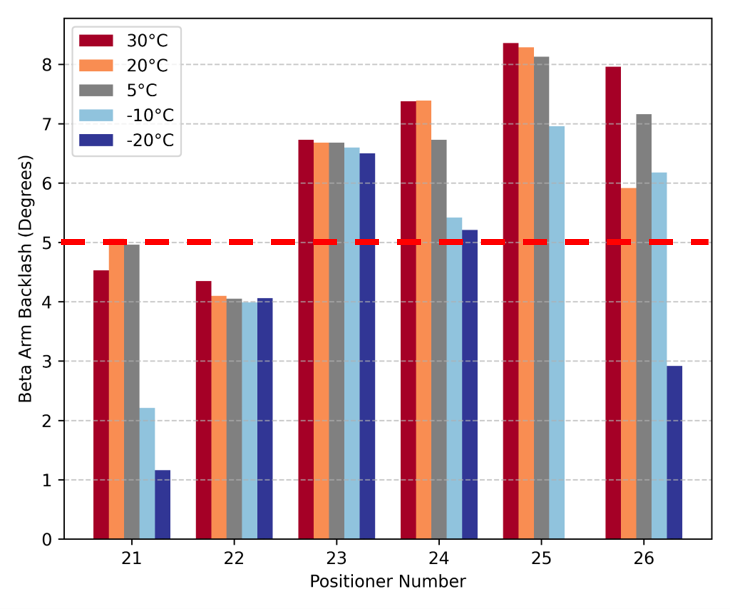}
         \caption{Beta arm}
         \label{fig:004_back_b_arm}
     \end{subfigure}
        \caption{Graphs showing the backlash for MPS6. The bar plots show the root-mean-square (RMS) values versus the positioner number (Specification: 5°)}
        \label{fig:004_back}
\end{figure}

The backlash of the alpha and beta arms is presented in Figure \ref{fig:004_back}, where it is shown that the alpha backlash for all positioners at all temperatures is below 7°. The beta backlash reaches more than 8° as observed in the figure. Although the specification is set at 5 deg for this positioner, the results remain encouraging in terms of possible software mitigation and mechanical improvement for later prototypes.

\subsection{Non-linearities}

In order to keep a lightweight presentation of the results we will present the two most interesting cases of non-linearities graphs obtained through the test campaign. The reader is welcome to contact the authors for more details.

\begin{comment}
\begin{figure}[H]
     \centering
     \begin{subfigure}[b]{0.49\textwidth}
        \centering
        \includegraphics[width=0.9\linewidth]{figures/004_results/NL_pos22_a_arm.png}
        \caption{Alpha arm}
        \label{fig:004_NL_pos22_a_arm}
     \end{subfigure}
     \hfill
     \begin{subfigure}[b]{0.49\textwidth}
         \centering
         \includegraphics[width=0.9\textwidth]{figures/004_results/NL_pos22_b_arm.png}
         \caption{Beta arm}
         \label{fig:004_NL_pos22_b_arm}
     \end{subfigure}
        \caption{Graphs showing the non-linearities for pos22 across the temperature range}
        \label{fig:004_NL_pos22}
\end{figure}
\end{comment}

\begin{figure}[H]
     \centering
     \begin{subfigure}[b]{0.49\textwidth}
        \centering
        \includegraphics[width=0.9\linewidth]{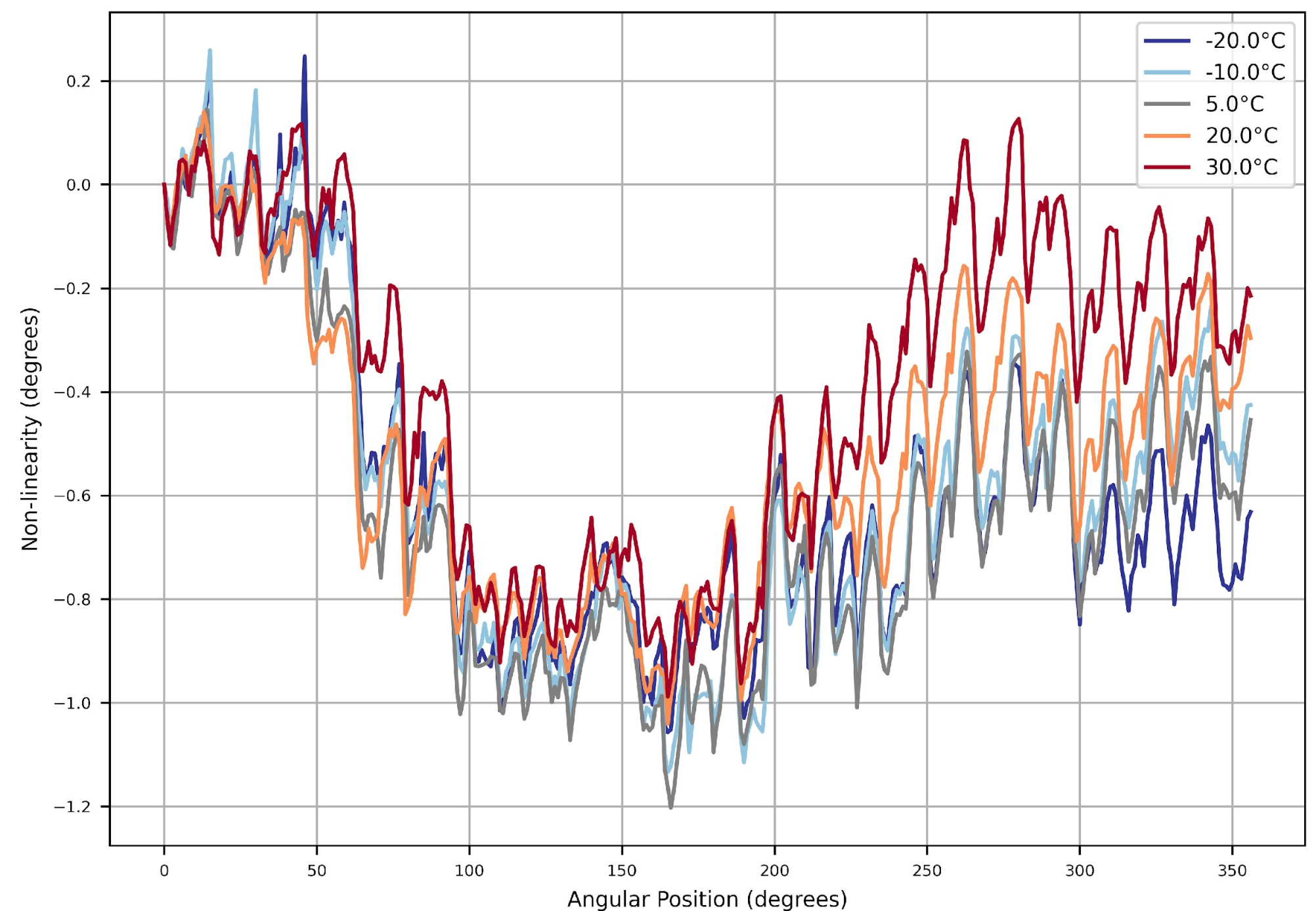}
        \caption{Alpha arm}
        \label{fig:004_NL_pos22_a_arm}
     \end{subfigure}
     \hfill
     \begin{subfigure}[b]{0.49\textwidth}
         \centering
         \includegraphics[width=0.9\textwidth]{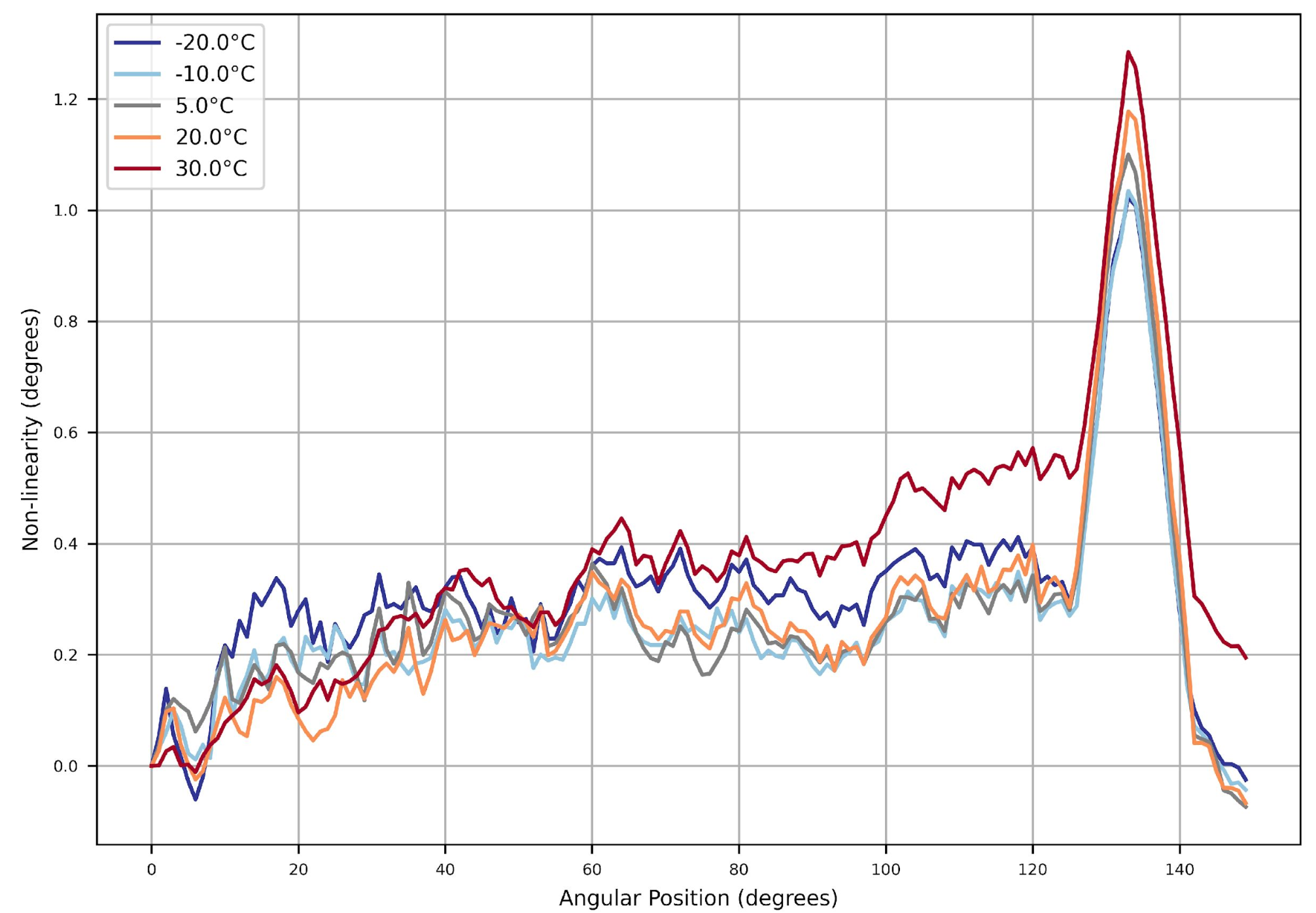}
         \caption{Beta arm}
         \label{fig:004_NL_pos22_b_arm}
     \end{subfigure}
        \caption{Graphs showing the non-linearities for pos22 across the temperature range}
        \label{fig:004_NL_pos22}
\end{figure}

\begin{comment}
\begin{figure}[H]
     \centering
     \begin{subfigure}[b]{0.49\textwidth}
        \centering
        \includegraphics[width=0.9\linewidth]{figures/004_results/NL_pos26_a_arm.png}
        \caption{Alpha arm}
        \label{fig:004_NL_pos26_a_arm}
     \end{subfigure}
     \hfill
     \begin{subfigure}[b]{0.49\textwidth}
         \centering
         \includegraphics[width=0.9\textwidth]{figures/004_results/NL_pos26_b_arm.png}
         \caption{Beta arm}
         \label{fig:004_NL_pos26_b_arm}
     \end{subfigure}
        \caption{Graphs showing the non-linearities for pos26 across the temperature range}
        \label{fig:004_NL_pos26}
\end{figure}
\end{comment}

\begin{figure}[H]
     \centering
     \begin{subfigure}[b]{0.49\textwidth}
        \centering
        \includegraphics[width=0.9\linewidth]{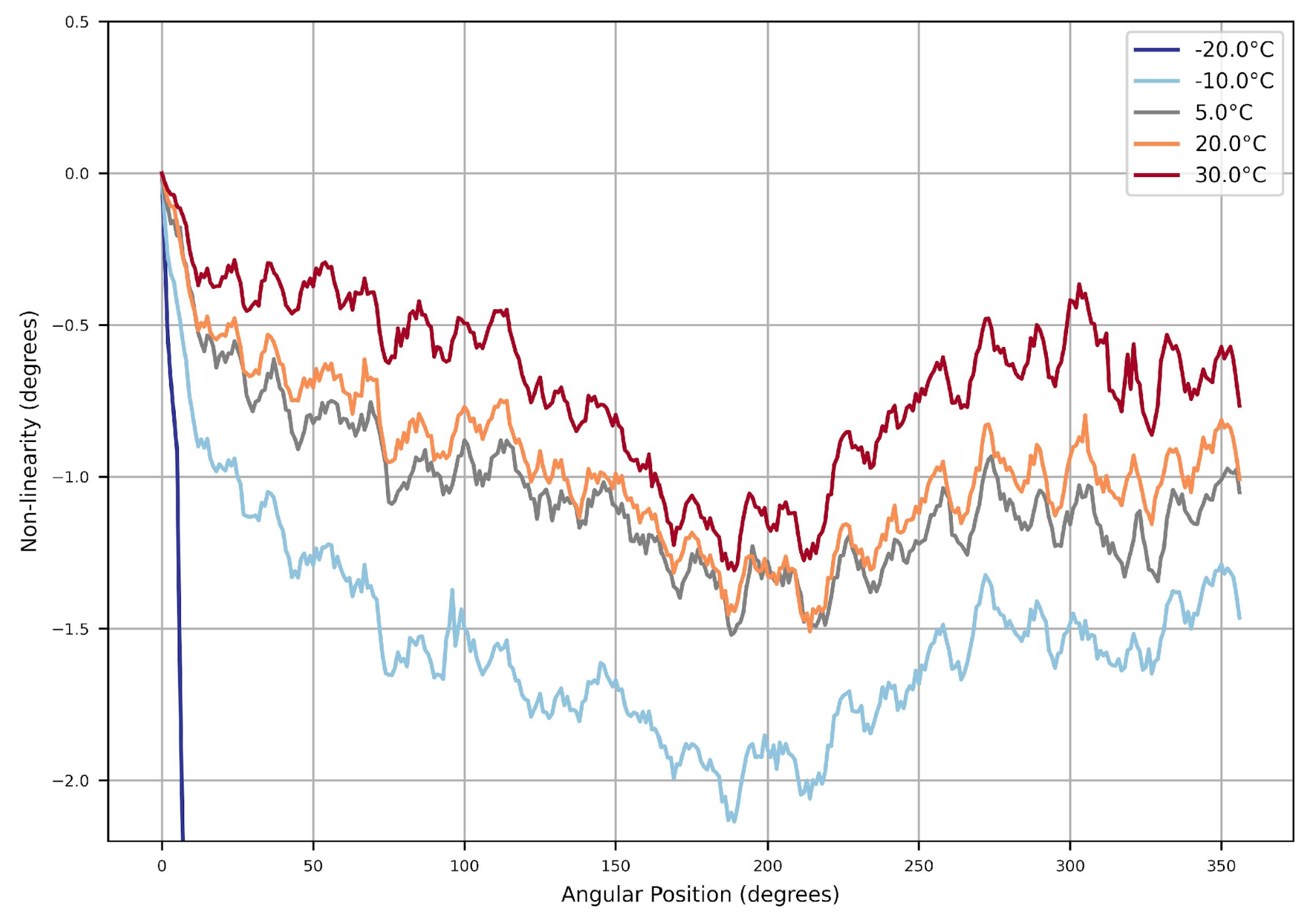}
        \caption{Alpha arm}
        \label{fig:004_NL_pos26_a_arm}
     \end{subfigure}
     \hfill
     \begin{subfigure}[b]{0.49\textwidth}
         \centering
         \includegraphics[width=0.94\textwidth]{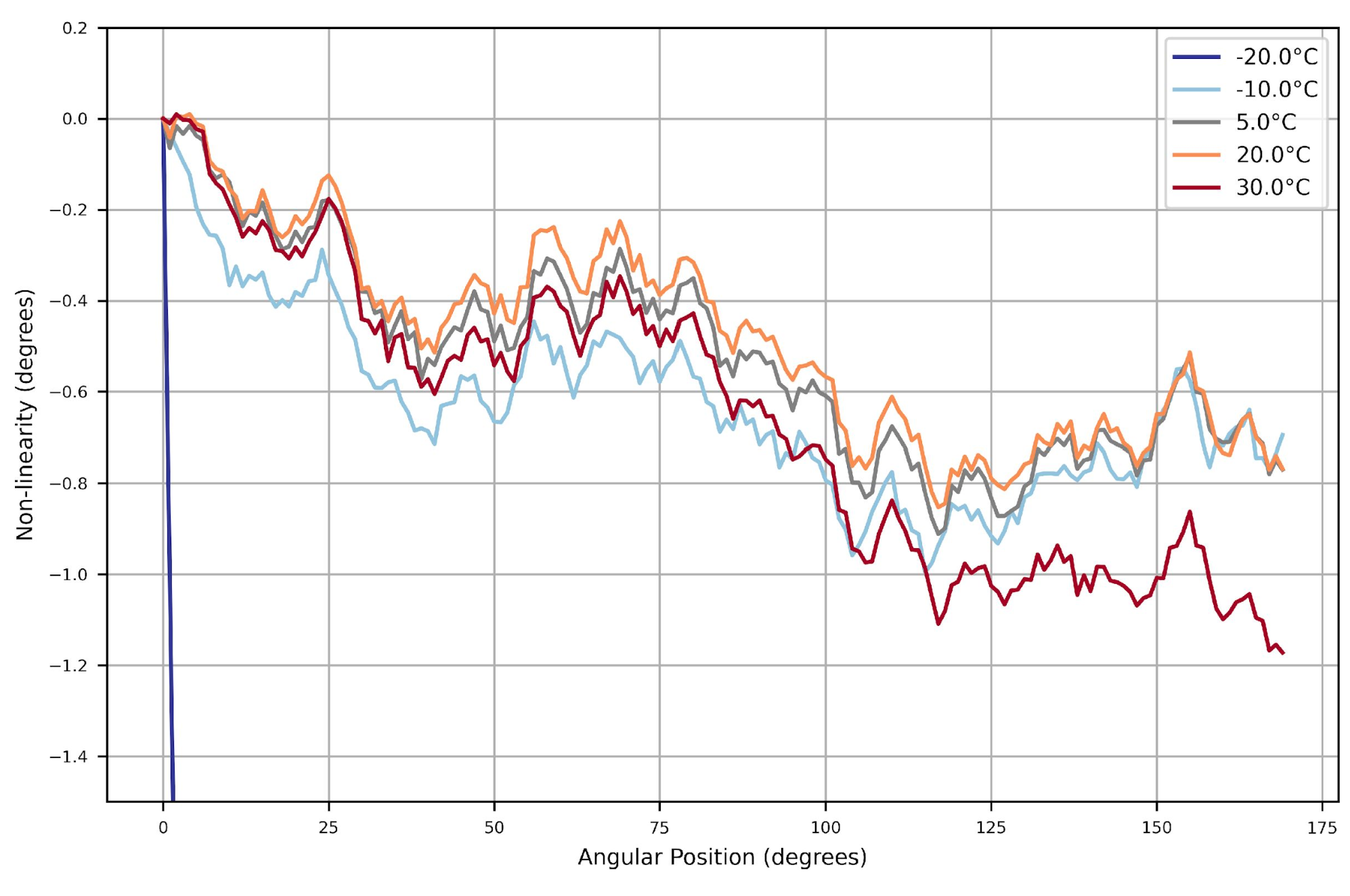}
         \caption{Beta arm}
         \label{fig:004_NL_pos26_b_arm}
     \end{subfigure}
        \caption{Graphs showing the non-linearities for pos26 across the temperature range}
        \label{fig:004_NL_pos26}
\end{figure}

Figures \ref{fig:004_NL_pos22} to \ref{fig:004_NL_pos26} present the non-linearity of pos22 and pos26 at the different temperatures.\\
Figure \ref{fig:004_NL_pos22_a_arm} and \ref{fig:004_NL_pos22_b_arm} show that the non-linearity curves seem to be mostly consistent over the temperature range for this positioner with very few interesting points such as the spike in \ref{fig:004_NL_pos22_b_arm} in the beta arm which may indicate a hard point at this angular location to be looked at.\\
Figure \ref{fig:004_NL_pos26_a_arm} and \ref{fig:004_NL_pos26_b_arm} show consistent results across the whole temperature range except at -20°C. This confirms the earlier suspicion about the positioner not moving at the particular temperature, because it does not execute the commanded sweep across the whole angular range.

%% file: 005_Conclusion.tex
\section{Discussion and Conclusion}

In this report, we presented the thermal testing of the first module prototype from MPS, the MPS6, for Stage-5 instrument survey. The experimental setup along with the results were thoroughly presented over a temperature range that goes from -20 degrees up to 30 °C. 

Sections \ref{sec:chamber_control} and \ref{sec:test_cycle} describe the setup and the testing methodology to have a better understanding of the experimental flow. It should be noted that to avoid introducing additional error due to the vibrations coming from the thermal chamber, the testing was conducted while the chamber was turned off. This means that the chamber was commanded to reach a particular temperature then shut down during each test cycle. Note that this first test campaign needed to be optimized for a tight test schedule and can therefore be improve through the following aspects for future module prototypes:

\begin{itemize}
    \item More thermal probes in contact with the module itself to directly assess its temperature instead of relying on the chamber probe
    \item More recent thermal chamber allowing for humidity control and less vibrations-induced during testing. The temperature could thus be actively controlled during test cycles which may also benefit from improvements
    \item Repeatability is only tested around a given angular position due to timing reason. Testing repeatability across more angular positions will increase even more the confidence in the system's reliability
\end{itemize}

\noindent Section \ref{sec:004_results} showed that the overall results of these positioners are very satisfactory for this first prototype. This encourages us to pursue prototyping in this direction with MPS Micro Precision Systems. \\
The highlighted inconsistencies are acceptable for this first prototyping stage and are being taken care of in the next round of prototypes. Later tests to determine the root cause of pos25 and pos26 issue at -20°C was identified to be the lubricant in the BLDC motors gearboxes which is now fixed with the motor manufacturers.